\documentclass[submission, Proceedings]{SciPost}

\hypersetup{
    colorlinks,
    linkcolor={red!50!black},
    citecolor={blue!50!black},
    urlcolor={blue!80!black}
}

\usepackage[bitstream-charter]{mathdesign}
\usepackage{physics}
\newcommand{\nn}{\nonumber\\}

\DeclareSymbolFont{usualmathcal}{OMS}{cmsy}{m}{n}
\DeclareSymbolFontAlphabet{\mathcal}{usualmathcal}

\begin{document}

\begin{center}{\Large \textbf{
Gluon pseudo-distributions at short distances\\
}}\end{center}

\begin{center}
I. Balitsky\textsuperscript{1,2},
W. Morris\textsuperscript{1,2} and
A. Radyushkin\textsuperscript{1,2$\star$}
\end{center}

\begin{center}
{\bf 1} Physics Department, Old Dominion University, Norfolk, VA 23529, USA
\\
{\bf 2} Thomas Jefferson National Accelerator Facility, Newport News, VA 23606, USA
\\
* radyush@jlab.org
\end{center}

\begin{center}
\today
\end{center}


\definecolor{palegray}{gray}{0.95}
\begin{center}
\colorbox{palegray}{
  \begin{tabular}{rr}
  \begin{minipage}{0.1\textwidth}
    \includegraphics[width=22mm]{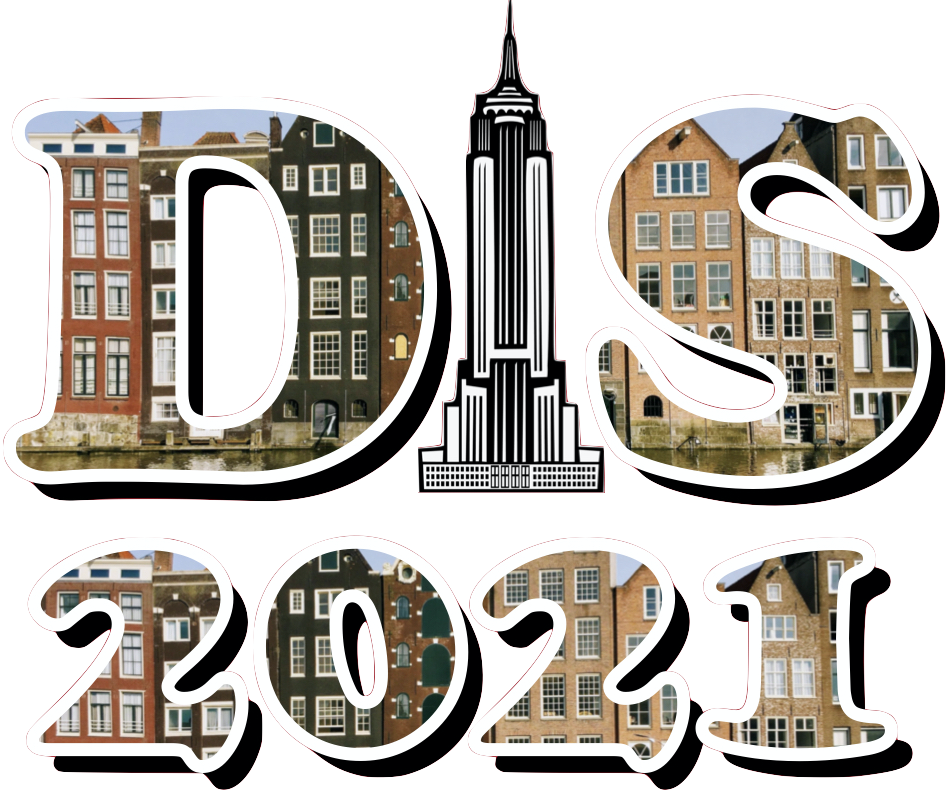}
  \end{minipage}
  &
  \begin{minipage}{0.75\textwidth}
    \begin{center}
    {\it Proceedings for the XXVIII International Workshop\\ on Deep-Inelastic Scattering and
Related Subjects,}\\
    {\it Stony Brook University, New York, USA, 12-16 April 2021} \\
    \doi{10.21468/SciPostPhysProc.?}\\
    \end{center}
  \end{minipage}
\end{tabular}
}
\end{center}

\section*{Abstract}
{\bf
We present the results that are necessary in the ongoing lattice calculations of the gluon parton distribution functions (PDFs) within the pseudo-PDF approach. We identify the two-gluon correlator functions that contain the invariant amplitude determining the gluon PDF in the light-cone $z^2 \to 0$ limit, and perform one-loop calculations in the coordinate representation in an explicitly gauge-invariant form. Ultraviolet (UV) terms, which contain $\ln (-z^2)$-dependence cancel in the reduced Ioffe-time distribution (ITD), and we obtain the matching relation between the reduced ITD and the light-cone ITD. Using a kernel form, we get a direct connection between lattice data for the reduced ITD and the normalized gluon PDF. 
}


\section{Introduction}
\label{sec:intro}

Lattice calculations of parton distribution functions (PDFs) are now a subject of considerable interest. 
Modern efforts aim at the extraction of PDFs $f(x)$ themselves rather than their $x^N$ moments. 
On the lattice, this may be achieved by switching from local operators to 
equal-time correlators \cite{Braun:2007wv,Ji:2013dva,Ma:2014jla,Radyushkin:2017cyf}.
 We use the pseudo-PDF approach   \cite{Radyushkin:2017cyf}, which 
is coordinate-space oriented, and  parton distributions are  extracted there by  taking the short-distance $z_3 \to 0$ limit. \\
Since the $z_3 \to 0$ limit is singular, one needs  matching relations to convert the Euclidean lattice data into the usual light-cone PDFs. Our  goal is to outline the pseudo-PDF approach to the extraction of unpolarized gluon PDFs \cite{Balitsky:2019krf}, and also to find one-loop matching conditions.\\
In the gluon case, the calculation is complicated by strict requirements of gauge invariance. In this situation, a very effective method is provided by the coordinate-representation approach of Ref. \cite{Balitsky:1987bk}. It is based on the background-field method and the heat-kernel expansion. It allows, starting with the original gauge-invariant bilocal operator, to find its modification by one-loop corrections. The results are obtained in an explicitly gauge-invariant form.

\section{Matrix elements}
%
%
%

The spin-averaged matrix element for operators comprised of two gluon fields with uncontracted indices is 
$
M_{\mu\alpha;\nu\beta} \left( z, p \right) \equiv \bra{p} G_{\mu\alpha} \left( z \right) \left[z,0\right] G_{\nu\beta} \left( 0 \right) \ket{p},
$
where $  \left[z,0\right] $ is the standard straight-line gauge link in the adjoint representation. 
Accounting for the antisymmetry of $G_{\rho\sigma}$ with respect to indices, and the available four-vectors $p$, and $z$, the decomposition of the matrix element into invariant amplitudes is:
\begin{align}
M_{\mu\alpha;\nu\beta}(z,p) &= \left(g_{\mu\nu} g_{\alpha\beta} -g_{\mu\beta} g_{\alpha\nu} \right)\mathcal{M}_{gg} (\nu, z^2)\nn
&\ +  \left( g_{\mu\nu} p_\alpha p_\beta - g_{\mu\beta} p_\alpha p_\nu - g_{\alpha\nu} p_\mu p_\beta + g_{\alpha\beta} p_\mu p_\nu \right) \mathcal{M}_{pp}(\nu, z^2) \nn
&\ + \left( g_{\mu\nu} z_\alpha z_\beta - g_{\mu\beta} z_\alpha z_\nu - g_{\alpha\nu} z_\mu z_\beta + g_{\alpha\beta} z_\mu z_\nu \right) \mathcal{M}_{zz}(\nu, z^2)   \nn
&\ + \left( g_{\mu\nu} z_\alpha p_\beta - g_{\mu\beta} z_\alpha p_\nu - g_{\alpha\nu} z_\mu p_\beta + g_{\alpha\beta} z_\mu p_\nu \right) \mathcal{M}_{zp} (\nu, z^2) \nn
&\ + \left( g_{\mu\nu} p_\alpha z_\beta - g_{\mu\beta} p_\alpha z_\nu - g_{\alpha\nu} p_\mu z_\beta + g_{\alpha\beta} p_\mu z_\nu \right) \mathcal{M}_{pz}(\nu, z^2)   \nn
&\ +  \left( p_\mu z_\alpha p_\nu z_\beta - p_\alpha z_\mu p_\nu z_\beta - p_\mu z_\alpha p_\beta z_\nu + p_\alpha z_\mu p_\beta z_\nu\right) \mathcal{M}_{ppzz} (\nu, z^2) ~,
\end{align}
where $\nu = - p\cdot z $ is the Ioffe time\cite{Ioffe:1969kf}. 
The light-cone distribution is obtained from \\ $g^{\alpha\beta} M_{+\alpha,+\beta} \left(z, p \right)$, where $z$ is taken to be in the ``minus'' direction $z = z_-$:
\begin{align}
g^{\alpha\beta} M_{+\alpha,\beta+} \left(z_-, p \right) = -2p_+^2 \mathcal{M}_{pp} \left( \nu, z^2 \right) ~ .
\end{align}
Thus the PDF is determined by $\mathcal{M}_{pp}$:
\begin{align}
-\mathcal{M}_{pp}\left( \nu, 0 \right) = {1 \over 2} \int_{-1}^1 \dd x e^{-ix\nu}x f_g\left( x\right) ~.
\end{align}
The procedure is then to take projections of the matrix element that contain the $\mathcal{M}_{pp}$ structure, and little of anything else. The projection that best meets this condition is $M_{0i;i0}$, whose decomposition is
$
M_{0i;i0} =   2 \mathcal{M}_{gg}  +2   p_0^2  \mathcal{M}_{pp} ~. ~
$
  We can remove the contaminating $\mathcal{M}_{gg}$ structure by adding the projection $M_{ij;ji}$, whose decomposition is
$
M_{ij;ji} = -2 \mathcal{M}_{gg} ~. \\
$
While all projections of the matrix element are individually multiplicatively renormalizable  \cite{Li:2018tpe}, they won't necessarily carry the same anomalous dimension; 
however, 
this addition works because both projections, $M_{0i;i0}$ and $M_{ij;ji}$, have the same anomalous dimension at one loop.
\section{One-loop corrections}
\subsection{Link self energy and ultraviolet divergences}
\label{sec:link}
%
The link self energy correction, given by
\begin{align}
 { g^2 N_c  \over 8\pi^2 } { \Gamma\left(1-\epsilon_{UV}\right) \left(-z^2\mu^2_{UV} + i\epsilon \right)^{\epsilon_{UV}} \over \left(2\epsilon_{UV} -1 \right)  \epsilon_{UV} }  G_{\mu\alpha}\left(z\right) G_{\nu\beta} \left(0\right)  \ ,
\end{align}
should, in principle, be zero on the light-cone. However, 
 this result contains linear ($\epsilon_{UV} = 1/2$) and logarithmic ($\epsilon_{UV} =0$) UV poles and a singularity on the light-cone $z^2=0$ after expansion in $\epsilon_{UV}$. 
In order to account for this, we explicitly separate the $z^2$ dependence generated by the UV singular terms and those in the QCD  
(or DGLAP ) evolution logarithms $\ln\left(-z^2\mu^2_{IR}\right)$.

\begin{figure}[h]
\centering
   \centerline{\includegraphics[height=2.05cm]{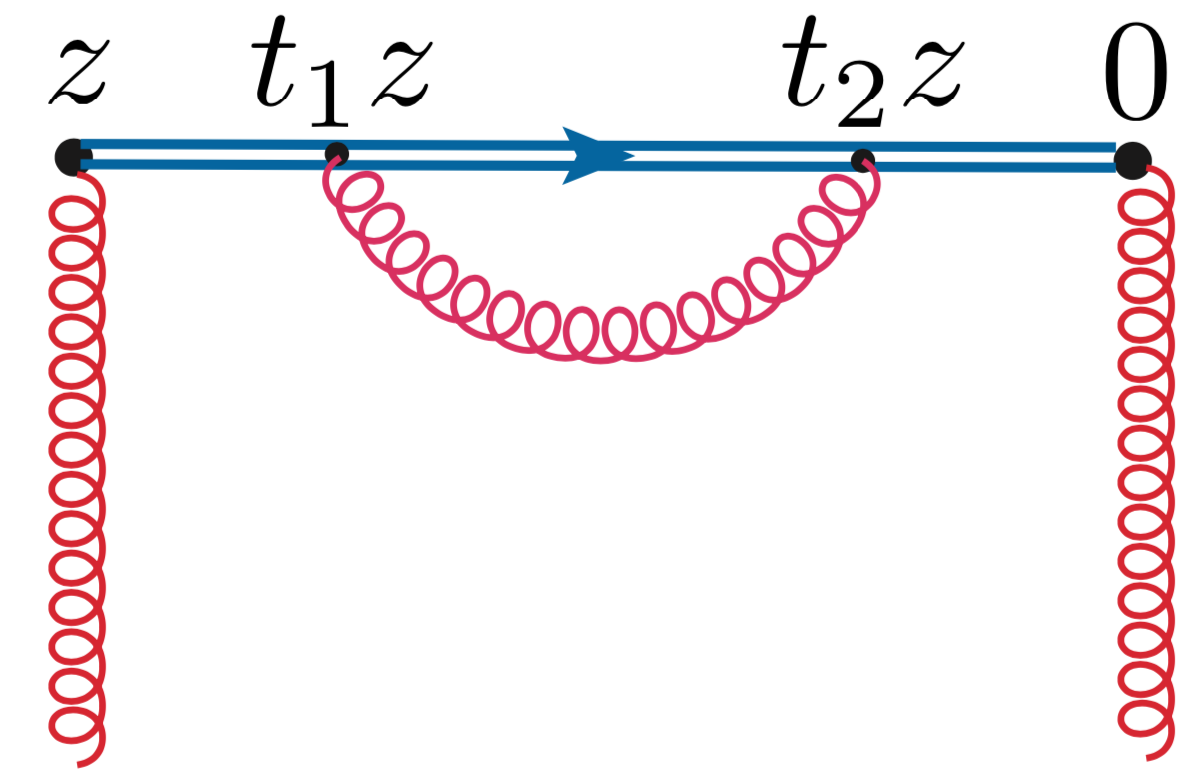}\hspace{5mm} \includegraphics[height=2.1 cm]{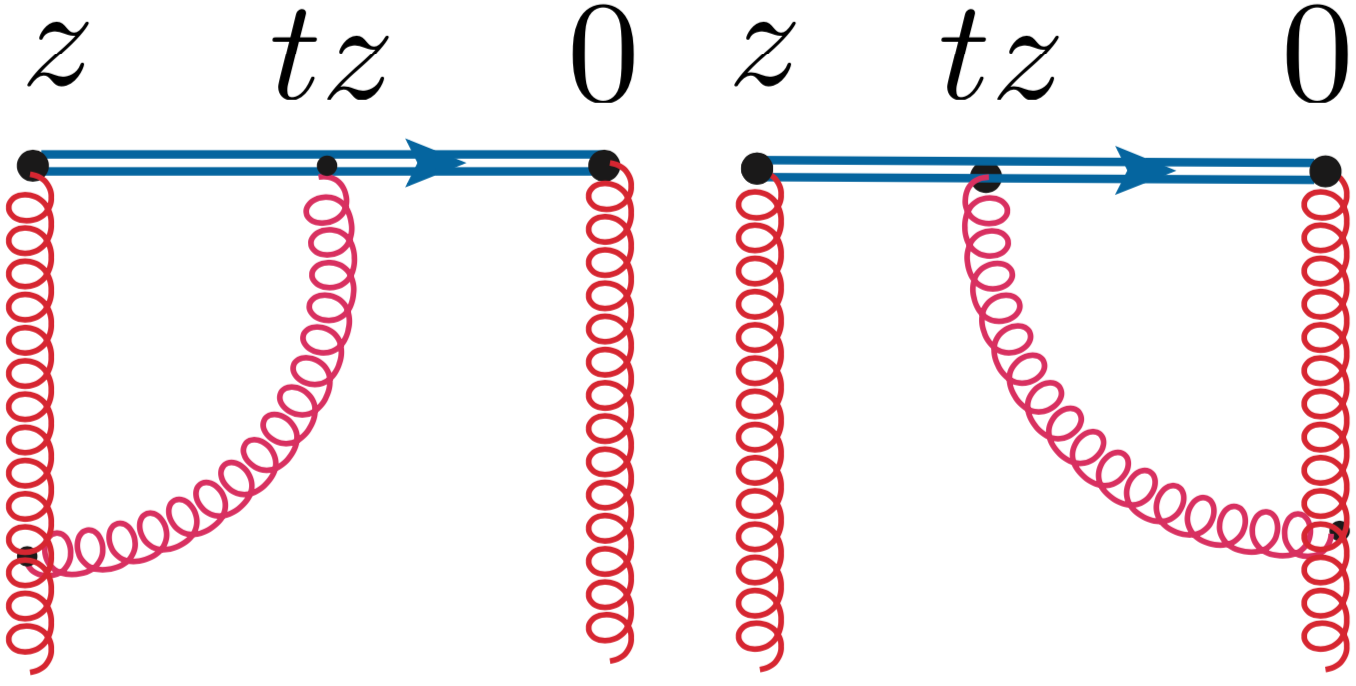}}
\caption{Self-energy-type correction for the gauge link and vertex diagrams with gluons coming out of the gauge link.}
\label{ref}
\end{figure}

\subsection{Vertex contributions}
When one  uses  the background-field technique, with gluon propagator in the  background-Feynman (bF) gauge \cite{Abbott:1980hw}, the three-gluon vertex differs from the usual Yang-Mills vertex. Therefore, the Feynman diagrams we use do not correspond one-to-one with the usual Feynman diagrams.
A consequence of this is, where one would have two linear UV divergences that cancel after addition of two diagrams, in our case that same divergence cancels implicitly. This can be seen from the evolution part of the vertex correction:
\begin{align}
{ g^2 N_c \over 8\pi^2 } { \Gamma\left( d/2 -2 \right) \over (d-3)\left( -z^2\right)^{d/2-2} } \int_0^1 \dd u \left[u^{3-d} -1\right]_+ G_{\mu\alpha} \left( \bar u z \right) G_{\nu\beta} \left( 0 \right)  \ ,
\end{align}
where 
the linear divergences present in the ``$u^{3-d}$'' part and the ``$-1$'' part cancel.\\
The full uncontracted vertex calculation contains also  a UV divergent and constant part, however this term is zero for the $M_{0i;i0}$ and the $M_{ij;ji}$ projections.

\subsection{Box and self-energy contributions}

The ``box'' correction is free of UV divergences, but gives a more complicated structure, generating a mixture of different operators corresponding to different projections of $G_{\mu\alpha}\left(\bar u z\right) G_{\nu\beta} \left(0 \right) $. \\
The full uncontracted result is too long for this paper, but it should be noted that the DGLAP part does not have the necessary plus-prescription form.
To get it, one should  add
 the contribution of the gluon self-energy diagrams. The relevant part of the result is the coefficient to the $\mathcal{M}_{pp}$ structure, $2\left(-u^3+u^2-2u+1 \right)$, which integrates to $1/6$. 
\begin{figure}[h]
\centering
   \centerline{\includegraphics[height=2cm]{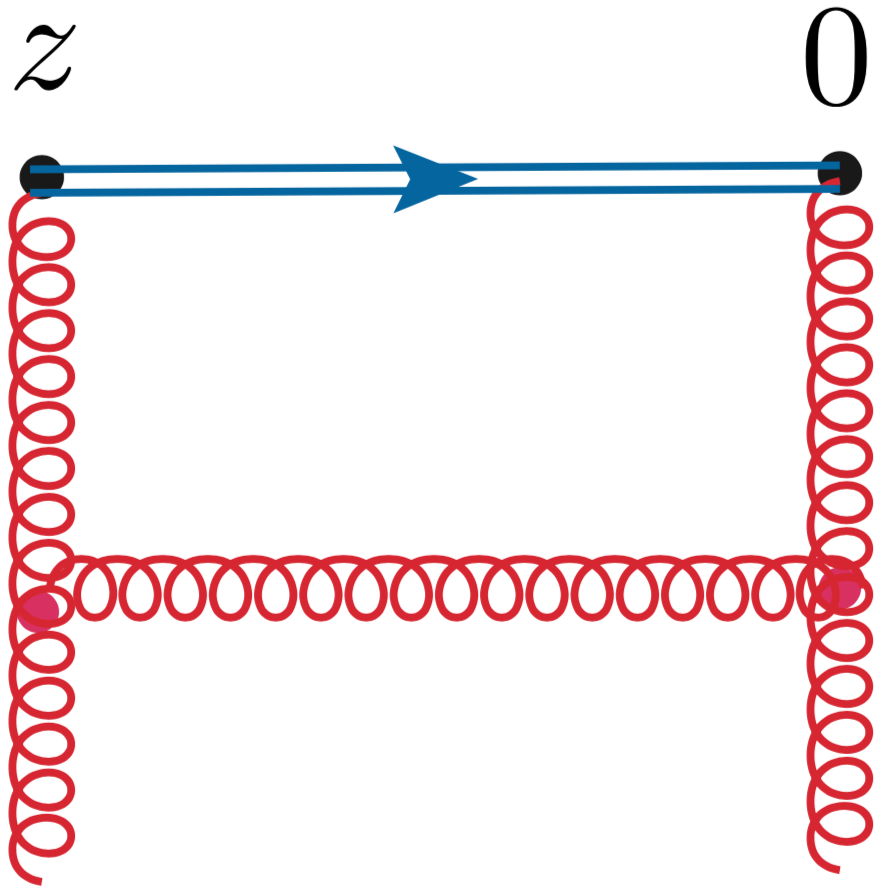} \hspace{5mm} \includegraphics[height=2.1cm]{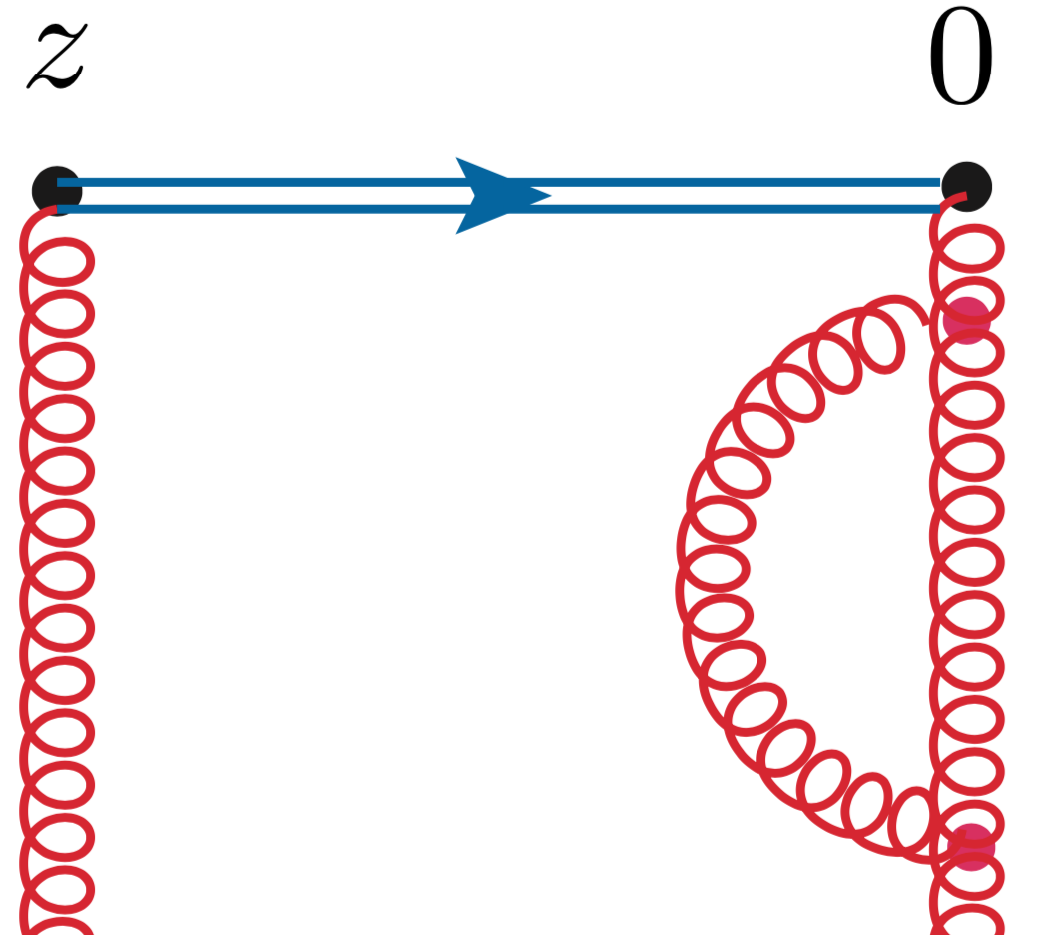} \hspace{5mm} \includegraphics[height=2.1cm]{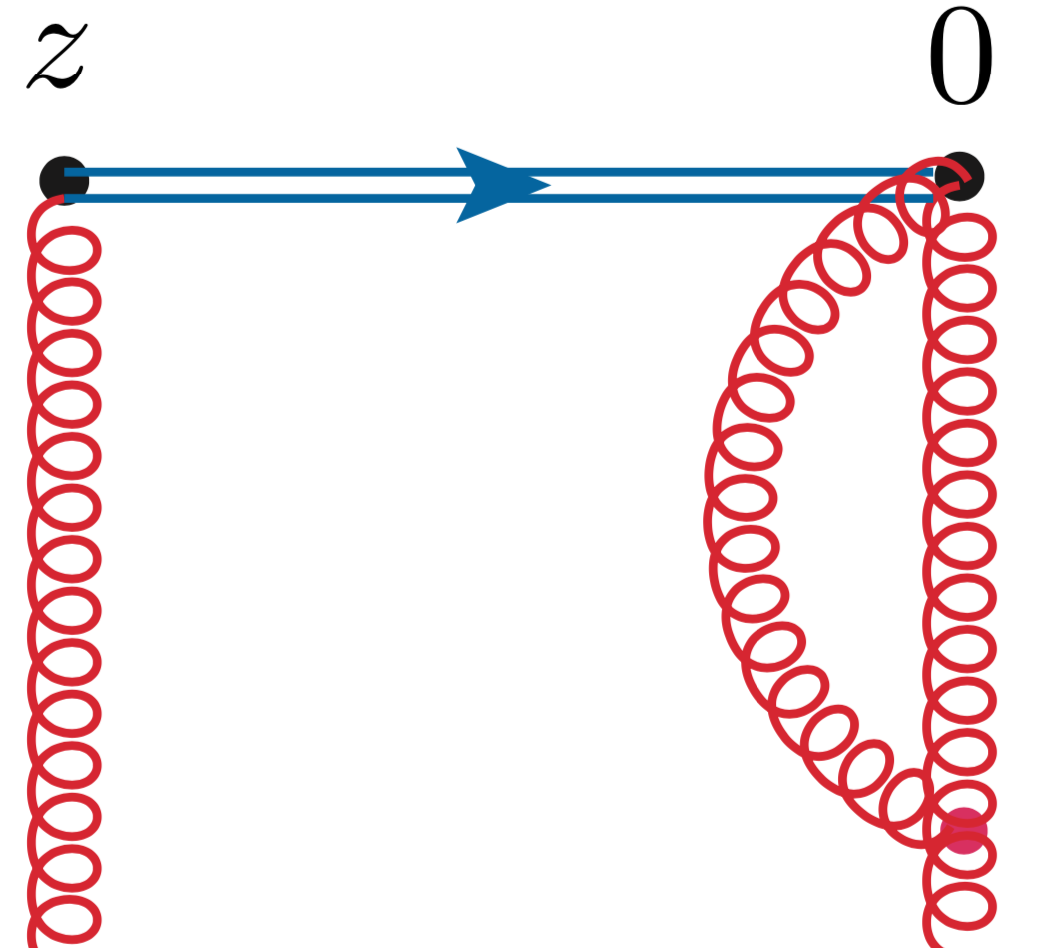}}
\caption{Box diagram  and gluon self-energy-type insertions into the right leg. }
\label{ref}
\end{figure}\\
The self-energy diagrams contain both UV and collinear divergences that generate logarithmic term $\ln\left(\mu^2_{IR} / \mu^2_{UV} \right)$. In order to obtain the necessary plus-prescription with the ``box'' diagram, one can separate this term into the difference $\ln\left( z_3^2 \mu^2_{IR} \right) - \ln\left( z_3^2 \mu^2_{UV} \right) $.  \\ 
The self-energy result is
\begin{align}
{ g^2 N_c \over 8\pi^2 } { 1 \over 2-d/2 } \left[2- { \beta_0 \over 2 N_c}  \right] G_{\mu\alpha} \left( z \right) G_{\nu\beta} \left( 0 \right) \ , 
\end{align}
where $\beta_0 = 11 N_c /3$ in gluodynamics, and ${ 1 \over 2-d/2 }$ is to be replaced by $\ln\left( z_3^2 \mu^2_{IR} \right) - \ln\left( z_3^2 \mu^2_{UV} \right) $. Substituting in the value of $\beta_0$, we get $1/6$, giving us the needed plus-prescription.

\section{DGLAP evolution structure}
\subsection{Reduced Ioffe-time distribution}
In order to  eliminate the link and self-energy UV divergences, we use the reduced ITD:
\begin{align}
\mathfrak{M} \left( \nu, z_3^2 \right) \equiv { \mathcal{M}_{pp} \left( \nu, z_3^2 \right) \over  \mathcal{M}_{pp} \left( 0, z_3^2 \right) } ~.
\end{align}
This method works because our operator is multiplicatively renormalizable, and because the DGLAP evolution logarithm  drops out of $\mathcal{M}_{pp} \left( 0, z_3^2 \right)$, so we're only and entirely removing the non-DGLAP related $z_3$ dependence.

\subsection{Matching relations}
Combining the one-loop gluon corrections, and the gluon-quark mixing term (that contains  
 the $gq$ evolution kernel 
$
\mathcal{B}_{gq}\left(u\right) = 1+ (1-u)^2 \, ) \ ,
$
\begin{figure}[h]
\centering
   \centerline{\includegraphics[height=2.2cm]{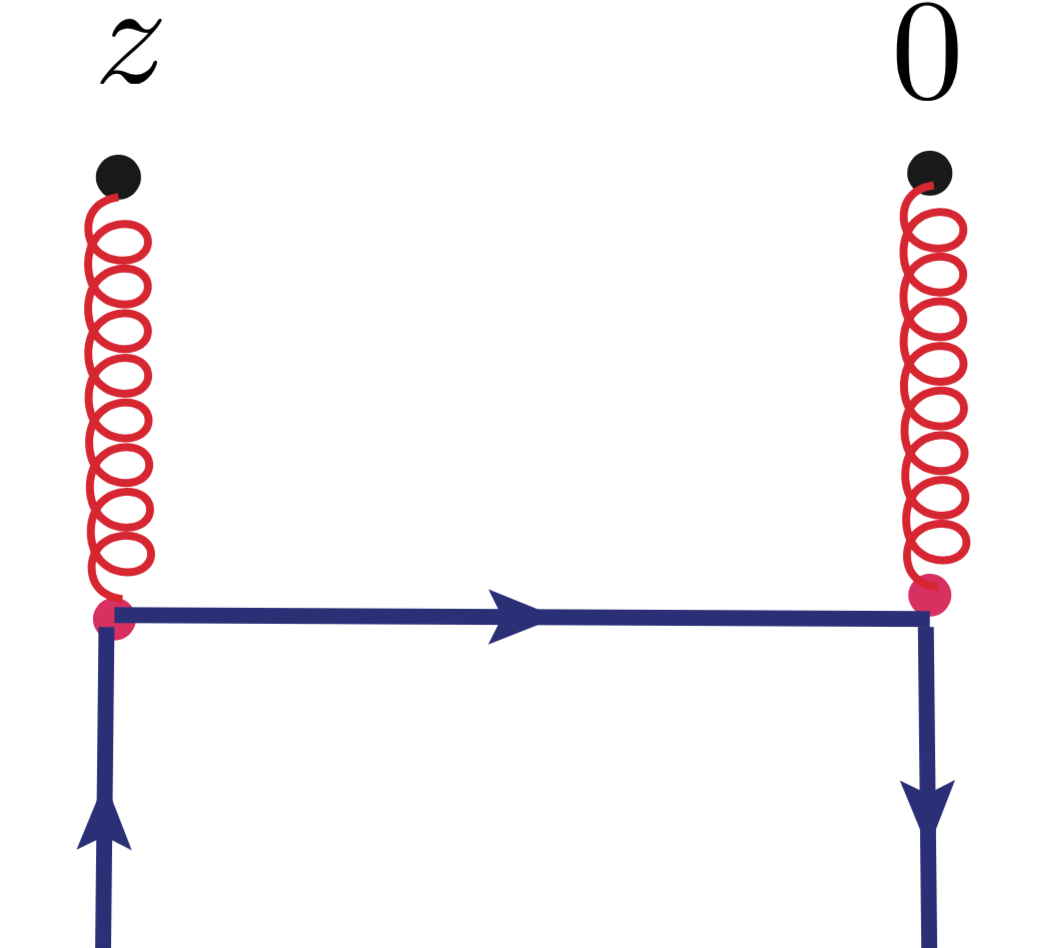}}
\caption{Gluon-quark mixing diagram.}
\label{ref}
\end{figure}
 we obtain the matching relation (excluding higher twist terms, or rather keeping only $\mathcal{M}_{pp}$):
\begin{align}
\mathfrak{M}(\nu,z_3^2) \mathcal{I}_g(0,\mu^2) &= \mathcal{I}_g(\nu,\mu^2) - { \alpha_s N_c \over 2\pi} \int_0^1 \dd u \mathcal{I}_g(u\nu,\mu^2) \left\{ \vphantom{1\over 1} \ln \left( z_3^2 \mu^2 e^{2\gamma_E} /4 \right) B_{gg} (u)  \right. \nn
&\ \left. +4 \left[ {u + \log (\bar u)  \over \bar u} \right]_+ + {2 \over 3} \left[ 1-u^3 \right]_+ \right\} \nn
&\ - {\alpha_s C_F \over 2\pi } \ln\left( z_3^2 \mu^2 e^{2\gamma_E} / 4 \right) \int_0^1 \dd w \left[\mathcal{I}_S(w\nu,\mu^2) - \mathcal{I}_S (0,\mu^2) \right] \mathcal{B}_{gq} (w) ~,
\end{align}
where $\mathfrak{M}(\nu,z_3^2)$, the reduced ITD, is our ``lattice function'', and $\mathcal{I}_g(\nu,\mu^2) $ and $\mathcal{I}_S(u\nu,\mu^2) $ are the light-cone ITDs. The gluon Altarelli-Parisi kernel is given by
\begin{align}
B_{gg} (u) = 2 \left[\frac{\left( 1-\bar uu\right)^2}{\bar u}\right]_+ \  . 
\end{align}
The gluon light-cone ITD can be directly related to the light-cone PDF through
\begin{align}
\mathcal{I}_g\left( \nu, \mu^2 \right) ={1 \over 2} \int_{-1}^1 \dd x e^{ix\nu} x f_g\left( x, \mu^2 \right) ~.
\end{align}
Because $x f_g\left( x, \mu^2 \right)$ is an even function of $x$, the real part of $\mathcal{I}_g\left( \nu, \mu^2 \right)$ is given by the cosine Fourier transform of $x f_g\left( x, \mu^2 \right)$, while the imaginary part vanishes. \\
 The factor $\mathcal{I}_g\left( 0, \mu^2 \right) = \expval{x}_{\mu^2}$ is the fraction of the hadron momentum carried by the gluons. It should be found from an independent lattice calculation (see, e.g. \cite{Yang:2018bft}).\\
The matching relation can be cast into a new kernel form in terms of the light-cone PDFs:
\begin{align}
\mathfrak{M}(\nu,z_3^2) = \int_0^1 \dd x { x f_g\left( x, \mu^2 \right) \over \expval{x}_{\mu^2} } R_{gg} \left( x\nu, z_3^2 \mu^2 \right) + \int_0^1 \dd x { x f_S\left( x, \mu^2 \right) \over \expval{x}_{\mu^2} } R_{gq} \left( x\nu, z_3^2 \mu^2 \right)~,
\end{align}
where 
\begin{align}
R_{gg} \left( y, z_3^2 \mu^2 \right) = \cos y - {\alpha_s N_c \over 2\pi }  \left\{\ln \left( z_3^2 \mu^2 e^{2\gamma_E}/4 \right) {R_B(y)} + {R_L(y)} + {R_C(y)}\right\}~,
\end{align}
and
\begin{align}
R_{gq} \left( y, z_3^2 \mu^2 \right) = - {\alpha_s N_c \over 2\pi } \ln\left( z_3^2 \mu^2 e^{2\gamma_E} / 4 \right) R_{\mathcal{B}}(y)
\end{align}
The various $R$ terms are given by cosine   transformations of the gluon kernel, log term, constant term, and mixing kernel, respectively, 
and are all perturbatively calculable expressions. \\
Using lattice data and  models  for $f_g (x, \mu^2) $ and $f_S (x, \mu^2 )$, one can fit their  parameters and $\alpha_s$.
\vspace{-9mm}

\section{Conclusion}
We presented the results of the calculations necessary in the ongoing work to extract gluon PDFs from the lattice using the method of pseudo-PDFs.
Specifically, we demonstrated 
$M_{0i;i0} + M_{ij;ji}$ to be the most promising combination 
of matrix elements
for obtaining the gluon PDF.  \\ 
We 
 gave the matching relations between the reduced pseudo-ITD and the light-cone ITD or light-cone PDF, demonstrating that lattice data and light-cone PDFs can be directly related.
\vspace{-5mm}

\section*{Acknowledgements}
We thank K. Orginos, J.-W. Qiu, D. Richards and S. Zhao for their 
 interest
 and 
discussions. 
\paragraph{Funding information}
This work is supported by Jefferson Science Associates, LLC under U.S. DOE Contract \#DE-AC05-06OR23177 and by U.S. DOE Grant \#DE-FG02-97ER41028.

\bibliography{references}

\nolinenumbers

\end{document}